%% file: samplepaper.tex
\documentclass[runningheads]{llncs}
\usepackage{graphicx}
\usepackage{color}
\usepackage{caption}
\usepackage{comment}
\usepackage{amsmath}
\usepackage{amssymb}
\usepackage{booktabs}
\usepackage{multirow}
\usepackage{hyperref}
\usepackage{epsfig,subfigure,multirow}
\DeclareMathOperator{\E}{\mathbb{E}}
\DeclareMathOperator{\argmin}{\text{argmin}}

\newcommand{\wbadd}[1]{{\color{black} #1}}
\newcommand{\wbrm}[1]{\wbadd{}}

\newcommand{\dgadd}[1]{{\color{black} #1}}
\newcommand{\dgrm}[1]{\dgadd{}}
\renewcommand{\vec}[1]{\mathbf{#1}}

\begin{document}
\title{Multi-agent Hierarchical Reinforcement Learning with Dynamic Termination}
%
%

\author{Dongge Han \and
Wendelin B\"ohmer \and
Michael Wooldridge \and Alex Rogers}
\authorrunning{D. Han et al.}
%
\institute{Dept. Computer Science, University of Oxford, Oxford, UK\\
\email{\{dongge.han, wendelin.boehmer, michael.wooldridge, alex.rogers\}@cs.ox.ac.uk}}

\maketitle              

\input{abstract.tex}

\keywords{Multi-agent Learning  \and Hierarchcial Reinforcement Learning }
\let\thefootnote\relax\footnote{* \scriptsize{Preprint presented at PRICAI 2019. The final authenticated version is available online at \url{https://doi.org/10.1007/978-3-030-29911-8_7}.}}
\input{body-text.tex}

%
%
%
 \bibliographystyle{splncs04}
 \bibliography{ijcai19}

\end{document}

%% file: abstract.tex
\begin{abstract}  
  In a multi-agent system, an agent's optimal policy will typically
  depend on the policies chosen by others. Therefore, a key issue in
  multi-agent systems research is that of predicting the behaviours of
  others, and responding promptly to changes in such behaviours.  One
  obvious possibility is for each agent to broadcast their current
  intention, for example, the currently executed option in a
  hierarchical reinforcement learning framework. However, this
  approach results in inflexibility of agents if options have an
  extended duration and are dynamic.
\noindent
While adjusting the executed option at each step improves flexibility
from a single-agent perspective, frequent changes in options can
induce inconsistency between an agent's actual behaviour and its
broadcast intention. In order to balance flexibility and
predictability, we propose a dynamic termination Bellman equation that
allows the agents to flexibly terminate their options. We evaluate our model empirically
on a set of multi-agent pursuit and taxi tasks, and show that our
agents learn to adapt flexibly across scenarios that require different
termination behaviours.
\end{abstract}

%% file: body-text.tex
\section{Introduction}
\input{intro}

\section{Related Work}
\input{related_work}

\section{Basic Definitions}
\input{background}

\section{Method}
\input{method}

\section{Experiments}
\input{experiments}

\section{Conclusions and Future Work}
\input{conclusion}

%% file: intro.tex
\begin{figure}[t]
\begin{center}
	\subfigure[Taxi A choosing a target]{\label{fig:scene1}\includegraphics[width=4.2cm]{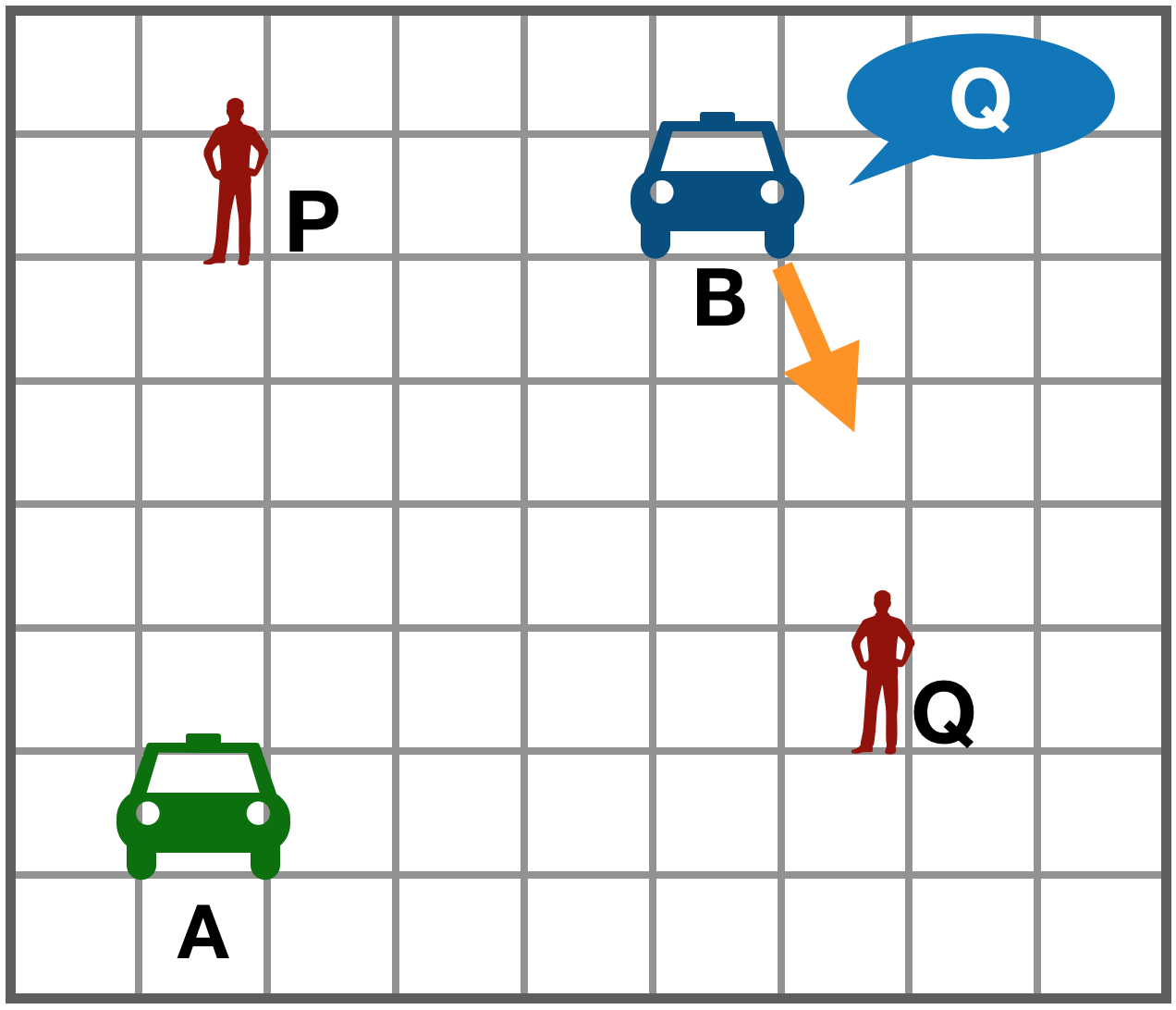}}
	\hspace{0.cm}
	\subfigure[Taxi B switching target]{\label{fig:scene2}\includegraphics[width=4.2cm]{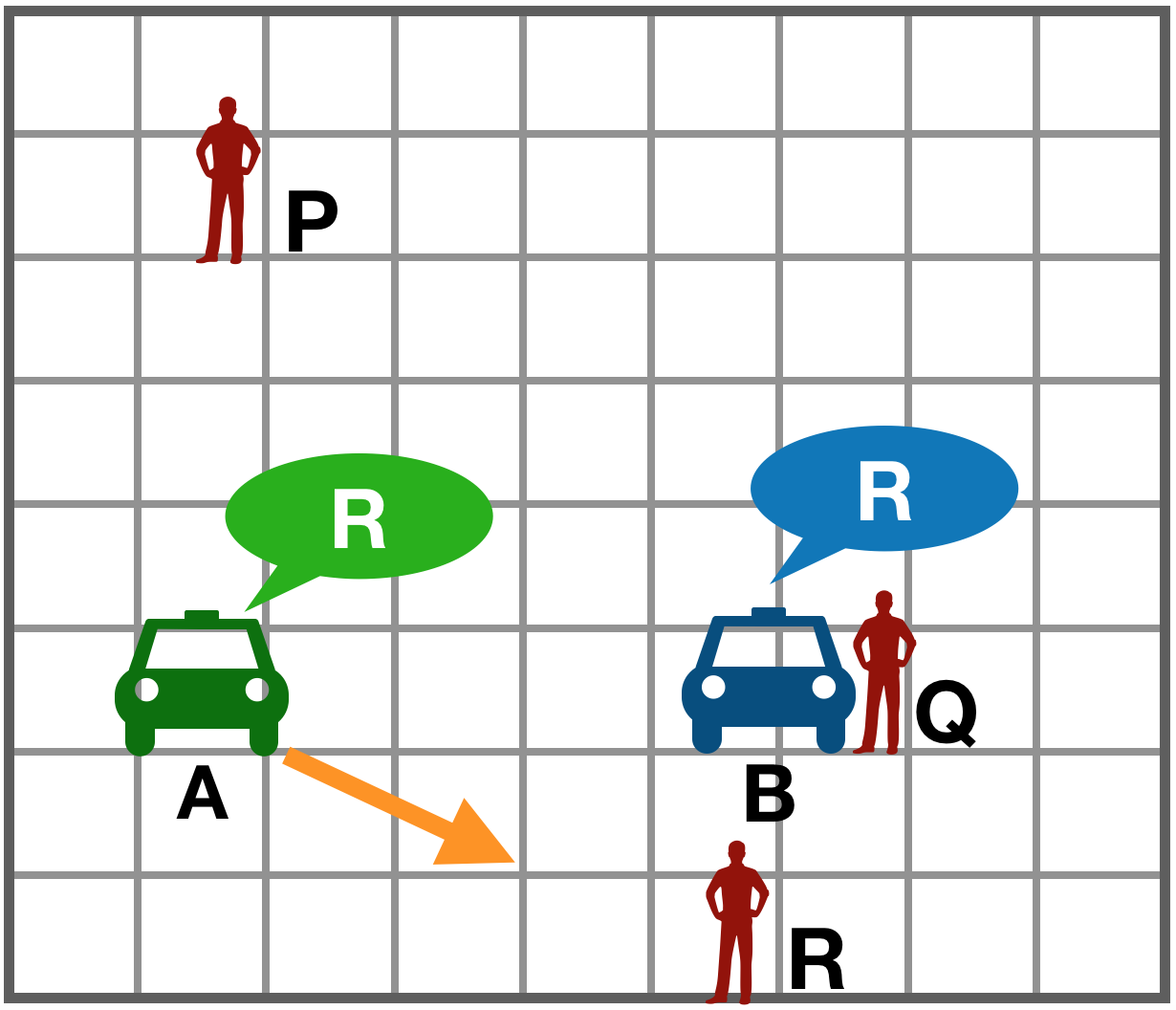}}
	\vspace{-0.4cm}
	\caption{Taxi Scenario Examples}
	\vspace{-0.8cm}
\end{center}
\label{fig:scenes}
\end{figure}

Many important real-world tasks are multi-agent by nature, such as
taxi coordination~\cite{lin2018efficient}, supply chain
management~\cite{giannakis2016multi}, and distributed
sensing~\cite{lesser2012distributed}. Despite the success of
single-agent reinforcement learning
(RL)~\cite{Sutton1998,mnih2015human}, multi-agent RL has remained as
an open problem. A challenge unique to multi-agent RL is that an
agent's optimal policy typically depends on the policies chosen by
others~\cite{stone2000multiagent}. Therefore, it is essential that an
agent takes into account the behaviours of others when choosing its
own actions.\dgrm{On the one hand, a centralized framework suffers
  from an exponentially growing joint action space. On the other hand,
  in decentralized frameworks, a significant challenge in an agent's
  perspective, is that the environment dynamics can be affected by the
  others' actions ~\cite{stone2000multiagent}, hence an optimal policy
  will depend on the policies of other agents. Therefore, it is
  essential that an agent learns the behaviours of others, and respond
  quickly to any changes in such behaviours.}
\noindent
One possible solution is to let each agent model and broadcast its intention, in order to indicate the agent's subsequent behaviours~\cite{foerster2016learning}. As an example, Figure \ref{fig:scene1} shows a taxi pickup scenario where taxi $A$ is choosing its next direction. Given the information that taxi $B$ is currently heading towards $Q$, taxi $A$ can determine passenger $P$ as its preferred option over $Q$.

\begin{figure}[!t]
\centering
    \includegraphics[width=0.8\columnwidth]{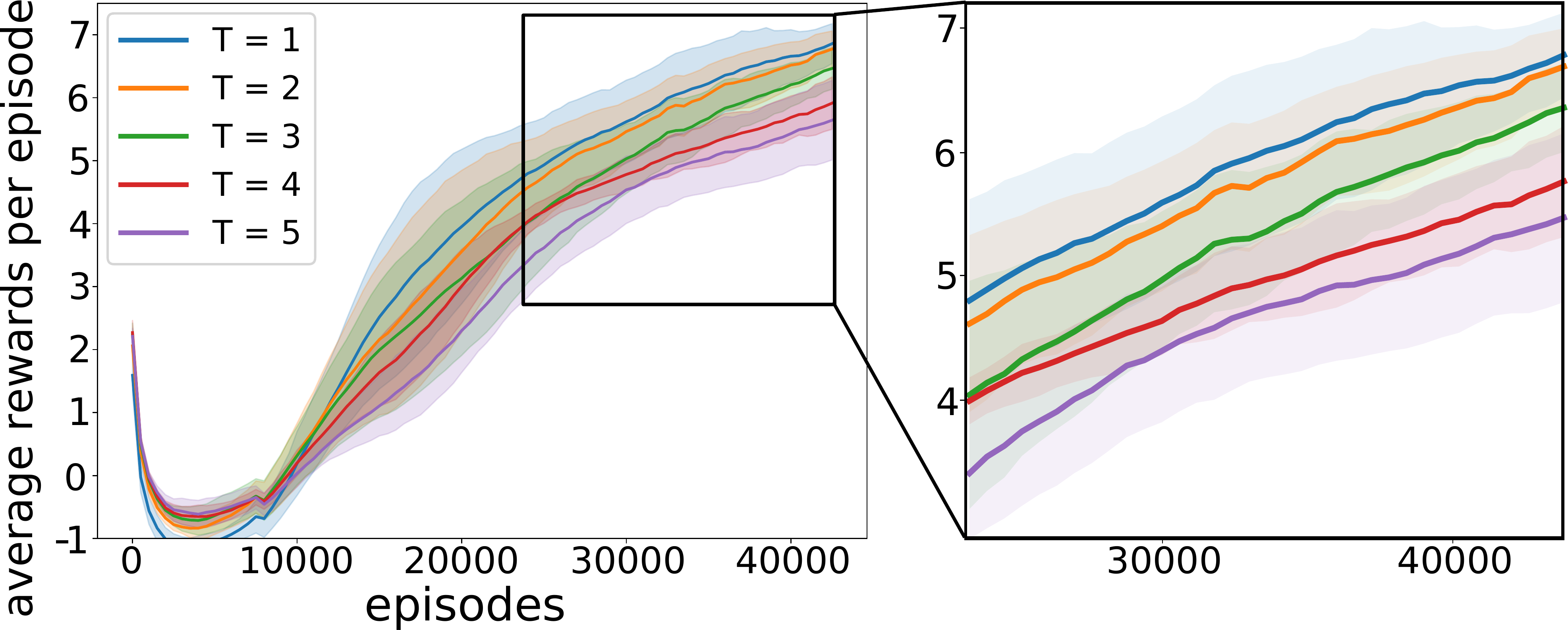}
    \caption{\label{fig:T} The effect of terminating options early, i.e., after T steps}
    \vspace{-0.6cm}
\end{figure}

Fortunately, hierarchical RL provides a simple solution for modeling
agents' intentions by allowing them to use \textit{options}, which are
subgoals that an agent aims to achieve in a finite
horizon. Makar et al. \cite{makar2001hierarchical}
proposed \emph{multi-agent hierarchical RL}, where hierarchical agents
broadcast their current options to the others.  However, despite the
advantage brought by using options, there can be a delay in an agent's
responses towards changes in the environment or others' behaviours,
due to the temporally-extended nature of options, which forbids the
agents from switching to another option before the current one is
terminated. In the scenario depicted in Figure \ref{fig:scene2}, while
taxi $A$ is going for passenger $R$, taxi $B$ finished picking up
passenger $Q$ and also switched towards $R$. In this case, taxi $A$
will miss the target $R$, but it cannot immediately switch its
target. \wbrm{Figure \ref{fig:T} shows a preliminary multi-agent taxi
  result where we interrupt the agents' current options after $T$
  timesteps. By reducing $T$, the agents gain higher flexibility for
  option switching, which also leads to increasing rewards.}

A potential solution to the delayed response challenge is to terminate options prematurely. 
\wbadd{Figure \ref{fig:T} shows the performance of a multi-agent taxi experiment where the agents' current options are interrupted after $T$ timesteps. By reducing $T$, the agents gain higher flexibility for option switching, which also leads to increasing rewards.}
This has been studied previously to address the problem of \textit{imperfect options} in single-agent settings, where an agent can improve its performance by
terminating and switching to an optimal option at each
step~\cite{sutton1999between}. 
However, this approach may no longer
prove advantageous in a multi-agent scenario. When an agent
frequently switches options, the broadcast option 
will be inconsistent with its subsequent behaviour. 
Consequently, the agent's behaviour becomes less predictable 
and the advantage of broadcasting options is diminished.

This poses a dilemma that is specific to multi-agent systems: 
excessive terminations makes an agent's behaviour unpredictable, 
while insufficient termination of options results 
in agents' inflexibility towards changes\cite{jennings1993commitments}. 
We will refer to an agent's \textit{flexibility} 
as the ability to switch options in response 
to changes in others or the environment. 
Furthermore, we will use \textit{predictability} to measure 
how far an agent will commit to its broadcast option. 
\noindent
In this paper, we propose an approach called \textit{dynamic
  termination}, which allows an agent to choose whether to terminate
its current option according to the state and others' options. This
approach balances flexibility and predictability, combining the
advantages of both.

An obvious approach to modelling dynamic termination is to use an
additional controller, which decides whether to terminate or to
continue with the current option at each step. In this paper, we
incorporate termination as an additional option for the high-level
controller. In this way, the Q-value of the newly introduced option is
associated consistently with the Q-values of the original options, and
our approach introduces negligible additional complexity to the
original model.
\noindent
We evaluate our model on the standard multi-agent pursuit and taxi coordination tasks across a range of parameters. The results
demonstrate that our dynamic termination model can significantly improve hierarchical multi-agent coordination and that it outperforms relevant state-of-the-art algorithms. The
contributions of our work are as follows:
\begin{enumerate}
	\item Based on the decentralized multi-agent options framework, we propose a novel dynamic termination scheme which allows an agent to flexibly terminate its current option. We show empirically that our model can greatly improve multi-agent coordination
	\item We propose a delayed communication method for an agent to approximate the joint Q-value. This method allows us to use intra-option learning, and reduces potentially costly communication
	\item We incorporate dynamic termination as an option to the high level controller network. This design introduces little additional model complexity, and allows us to represent the termination of all options in a consistent manner
\end{enumerate}

\noindent
In addition, we adopted several methods that benefits the model architecture
	and training: deep Q networks and parameter sharing
	reduce state space and model complexity; 
	adapting intra-option learning~\cite{sutton1999between} 
	to multiple agents yields better sample efficiency; and
	an off-policy training scheme~\cite{harutyunyan2017learning} 
	for exploration.
	
\dgrm{The remainder of the paper is structured as follows: related work and
background are described in Section 2 and 3, respectively. In Section
4, we introduce the multi-agent options framework and our proposed
dynamic termination operator. Empirical results are presented in
Section 5. We conclude and discuss possible future work in Section 6.}

%% file: related_work.tex
Makar et al.\cite{makar2001hierarchical}
appear to have been the first to combine multi-agent and hierarchical
RL, through the MaxQ framework~\cite{dietterich2000hierarchical}. We
build on their work, with the following changes: First, the use of
tabular Q-learning is insufficient for large state spaces. Therefore,
we adopt deep Q networks for parameterizing state and action
spaces. Second, we adapt intra-option learning to multi-agent
systems~\cite{sutton1999between}, which greatly improves the sample
efficiency. Third, we adopt a delayed communication channel to prevent
costly communication, and joint optimization.  And finally, as options
cannot be terminated before their predefined termination condition,
tasks are limited to the use of perfect options and agents experience
the delayed response problem.

Our solution to the delayed response problem is related to works on
interrupting imperfect options, i.e., when the set of available
options are not perfectly suited to the task, an agent can choose to
terminate its options dynamically in order to improve its
performance. Sutton et al.\cite{sutton1999between}
introduced a mechanism for interrupting options whenever a better
option appears, and Harutyunyan et al.\cite{harutyunyan2017learning}
proposed a termination framework which improves upon this idea with
better exploration. This is achieved by off-policy learning, which
uses an additional behaviour policy for longer options.

Bacon et al.\cite{bacon2017option} proposed a
dynamically terminating model for their Option-critic framework. In
comparison, we use the Q-learning framework instead of policy
gradient; and we focus on addressing the coordination problems in a
multi-agent system. Moreover, our Q-value for dynamic termination does
not depend on the currently executing option, which significantly
reduces the model complexity and also improve sample efficiency due to
off-policy training, i.e., the value of terminating can be learned
with any executed option.

In the multi-agent learning literature,
Riedmiller et al.\cite{riedmiller2005mutlioption}
proposed the multi-option framework. This is a centralized model in
which multiple agents are considered as a single meta-agent that
chooses a joint option $\vec{o} =(o_1, \ldots, o_n)$. In contrast, our
model uses a decentralized scheme where each agent $i$ chooses and
executes its own option $o_i$. This reduces the action space of the
high level controller from $|O|^n$ to $|O|$, where $O$ is the set of
all options, and $n$ is the number of agents.

Our model also draws upon the independent Q-learning framework
proposed by Tan\cite{Tan1997}, where each agent
independently learns its own policy on primitive actions, while
treating other agents as part of the environment. Additionally in our
model, each agent conditions on the others' broadcast options as part
of its observation when choosing the next option. We will discuss the
detailed formulation in section \ref{sec:delayed}.

%% file: background.tex


We first introduce the essential concepts in reinforcement learning
(RL), followed by multi-agent RL, hierarchical RL, intra-option
learning and off-policy termination.

A \textit{Markov Decision Process}~\cite{Sutton1998} is given by a
tuple $\langle \mathcal S, \mathcal A, R, P, \gamma \rangle$, where
$\mathcal S$ denotes a set of states, $\mathcal A$ a set of actions,
$P$ the stationary transition probability $P(s_{t+1}|s_t, a_t)$ from
state $s_t$ to state $s_{t+1}$ after executing action $a_t$, $R$ is
the average reward function $r_t := R(s_t, a_t)$, and
$\gamma \in [0, 1)$ is the discount factor.  A policy $\pi(a_t|s_t)$
is a distribution over actions $a_t$ given the state $s_t$.  The
objective of a RL agent is to learn an optimal policy $\pi^*$, which
maximizes the expected cumulative discounted future rewards.  The {\em
  Q-value} of the optimal policy conditions this return on an action
$a_t$ that has been selected in a state $s_t$: \vspace{-0mm}
\begin{equation} \label{eq:q_value} \footnotesize
    Q^*(s_t, a_t) 
    = \E\!\Big[{\textstyle\sum\limits_{\tau=0}^\infty} \, 
        \gamma^\tau r_{t+\tau}\Big]
    = r_t + \gamma \max_{a'} \E\!\big[ Q^*(s_{t+1}, a')\big] .
    \hspace{-1mm}
\end{equation}
Q-learning learns the Q-value of the optimal policy by interacting
with a discrete environment~\cite{Watkins1992}.  Continuous and
high-dimensional states require function
approximation~\cite{Sutton1998}, for example deep convolutional neural
networks (DQN)~\cite{Mnih2013,mnih2015human}.  To improve the
stability of gradient decent, DQN introduces an {\em experience replay
  buffer} to store transitions that have already been seen.  Each
update step samples a batch of past transitions and minimizes the
mean-squared error between the left and right side of Equation
\ref{eq:q_value}.

In multi-agent reinforcement learning, $n$ agents interact with the
same environment.  The major difference to the single agent case is
that the joint action space
$\mathcal A = \mathcal A^1 \times \cdots \times \mathcal A^n$ of all
agents grows exponential in $n$.  Independent Q-learning addresses
this by {\em decentralizing} decisions~\cite{Tan1997}: each agent
learns a Q-value function that is independent of the actions of all
other agents.  This treats others as part of the environment and can
lead to unstable DQN learning~\cite{foerster2017-stabilising}.  Other
approaches combine decentralized functions with a learned centralized
network~\cite{rashid2018qmix} or train decentralized Actor-Critic
architectures with centralized
baselines~\cite{foerster2017counterfactual}.

We now describe some important concepts related to hierarchical
reinforcement learning (HRL).  The Options
Framework~\cite{sutton1999between} is one of the most common HRL
frameworks, which defines a two-level hierarchy, and introduces
options as temporally extended actions.  Options $o$ are defined as
triples $\langle\mathcal{I}^o, \beta^o, \pi^o \rangle$, where
$\mathcal{I}^o \subseteq \mathcal{S}$ is the initiation set and
$\beta^o: \mathcal{S} \rightarrow [0,1]$ is the option termination
condition.  $\pi^o: \mathcal{S} \rightarrow \mathcal{A}$ is a
deterministic option policy that selects primitive actions to achieve
the target of the option.  On reaching the termination condition in
state $s'$, an agent can select a new option from the set
$\mathcal O(s') := \{ o \,|\, s' \in \mathcal{I}^o\}$.  A Semi-Markov
Decision Process (SMDP)~\cite{sutton1999between} defines the optimal
Q-value:

\vspace{-4mm}
\begin{equation} \label{eq:options}
    Q(s_t, o_t) \;\;=\;\; \E\!\Big[{\textstyle \sum\limits_{\tau=0}^{k-1}}
        \gamma^\tau r_{t+\tau} +\, \gamma^k 
        \kern-2ex\max_{o' \in \mathcal{O}(s_{t+k})}\kern-2ex 
        Q(s_{t+k}, o')\Big] \,,
\end{equation}
where $k$ refers to the number of steps until the termination
condition $\beta^{o_t}(s_{t+k}) = 1$ is fulfilled.

\dgrm{Intra-option Learning
~\cite{sutton1999between, sutton1998intra} 
is designed as an off-policy training method for options to improve the sample efficiency. 
In the original Options framework, 
the value of choosing an option is updated 
given by Equation \ref{eq:options}. 
During training, 
the value of an option $o$ is updated only 
when it is executed till termination 
and reaching the new state $s_{t+k}$. }
To improve sample efficiency, 
Intra-option Learning~\cite{sutton1999between,sutton1998intra}  was proposed as an off-policy learning method which at each time step $t$ updates all options 
that are in agreement with the executed action,
i.e.~$\forall o \in \{o \,|\, \pi^o(s_t) = a_t\}$ holds:
\begin{eqnarray}\label{eq:intra_option}
    Q(s_t, o) &=& r_t + \gamma 
        \E\!\big[ U(s_{t+1}, o) \big] \\
    \nonumber
    U(s, o) &=& \big(1 - \beta^o(s)\big) \, Q(s, o) \;+\, \beta^o(s) 
        \kern-.5ex\max_{o' \in \mathcal{O}(s)}\kern-1ex Q(s, o') .
\end{eqnarray}
Here $U(s_{t+1},o)$ is the TD-target ~\cite{tesauro1995temporal}: if
$o$ is terminating in the next state, the TD-target will be the value
of choosing the next optimal option.  If not, the target will be the
value of continuing with option $o$.
\noindent
Updating multiple options vastly improves the efficiency of training. 
Consider a grid-world navigation case where an agent is going for some goal location, 
and each coordinate corresponds to the sub-goal of an option. 
When the agent takes a primitive action $a_t^j$ and reaches the next position, 
all options $o^j$ that would have chosen that action will be updated.
Each transition updates therefore 
a significant fraction of the options,
which massively improves sample efficiency.

\dgrm{
\noindent
The \textit{MAXQ framework} ~\cite{dietterich2000hierarchical} is another HRL variant which 
defines a value decomposition scheme for 
a single-agent, multi-level task hierarchy which decomposes 
a task into sub-tasks recursively down to primitive actions. For the purpose of this paper, 
we will only focus on describing the two-level MAXQ framework. 
Each option $o$ is defined as 
$\langle \mathcal{S}^o,\mathcal{I}^o, \beta^o, O^o, \pi^o \rangle$, 
where $\mathcal S^o$ is the state space, 
$\mathcal{I}^o$ is the initiation set, 
$\beta^o$ is the termination condition, 
$O^o$ is the set of available actions of 
the next lower layer to achieve $o$, 
which are in this case primitive actions.

Makar et al.~\shortcite{makar2001hierarchical} extended the MAXQ framework 
to \textit{Multi-agent MAXQ framework}, 
where each agent learns its own Q-value $Q^j$. 
They define \textit{cooperative sub-tasks} where the Q-value 
of an agent is conditioned on the others' current sub-tasks. 
Under state $s$ and parent task $m$, 
the value of agent $j$ choosing sub-task $o^j$, 
while others are executing 
$\vec o^{-j} := (o^1, \ldots, o^{j-1}, o^{j+1}, \ldots, o^n)$, 
is decomposed as the expected return while $j$ is executing $o_j$, 
and the expected future return after $o_j$ is finished:
\begin{equation} \label{eq:multiagentQ}
    Q^j(m, s_{t}, \vec o^{-j}_{t}, o^j_{t}) 
    = \E\!\Big[{\textstyle\sum\limits_{\tau=0}^{k-1}} r_t
    + \max_{o'^j} Q^j(m, s_{t+k}, \vec o^{-j}_{t+k}, o'^j) \Big] \,. 
\end{equation}
This model exhibits two problems:
(i) if two agents finish their subtask simultaneously,
they would be forced to perform a joint optimization of 
their decentralized Q-values to make an optimal decision.
(ii) the current subtask of agent $j$ may no longer be 
the optimal choice as soon as any other agent finishes their task.
Optimal decision making would therefore 
require all agents to stop their current tasks
and jointly re-distribute the subtasks
every time an agent finishes its respective task.\\}

\dgrm{In the next section we will derive a method that addresses 
the problems of Makar et al.~\shortcite{makar2001hierarchical}
and involves a dynamic termination operator.
The idea is inspired by \textit{Option Termination} in single-agent learning 
that addresses the problem of imperfect options.}

As introduced by
Sutton et al.~\cite{sutton1999between}, when the
set of available options are not suited to the task, an agent can
improve its performance by terminating at each step and switching to
an optimal
option. Harutyunyan et al.~\cite{harutyunyan2017learning}
have shown that this approach improves the agent's performance
significantly, but has an adverse effect on exploration: temporally
extended options can explore the state space more consistently, which
is lost by early termination.  The authors therefore advocate the use
of intra-option learning to update the Q-value off-policy, while
executing a different exploration policy that follows a selected
option for multiple steps before terminating.

%% file: method.tex
In this section we will present our framework for deep decentralized
hierarchical multi-agent Q-learning.  Our model uses delayed
communication to approximate the decisions of a centralized {\em joint
  policy}, which avoids many problems usually associated with joint
optimization.  This induces new challenges such as a {\em delayed
  response} of agents, and requires us to define a novel {\em dynamic
  termination update equation}.

\begin{figure*}[t] 
    \vspace{-4mm}
    \centering
    \includegraphics[width=13cm, height=4.2cm]{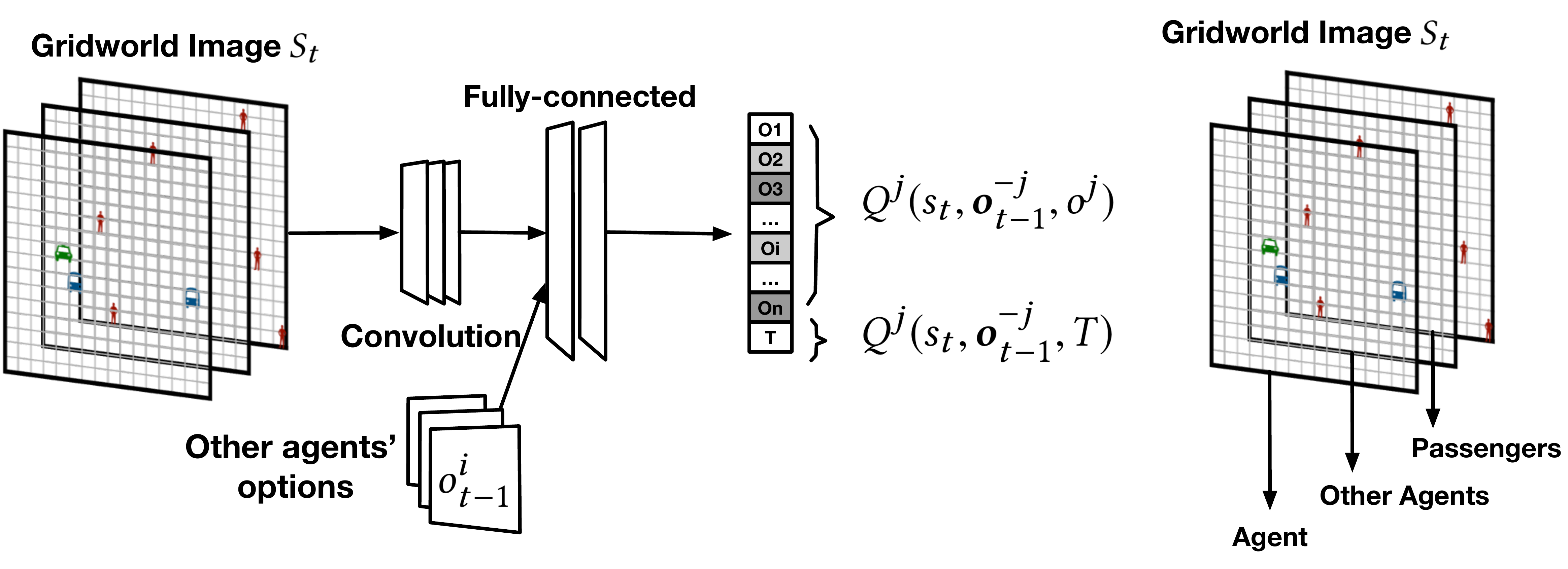}
    \vspace{-0.4cm}
    \caption{Dynamic termination Q-value network architecture.}
    \label{fig:dyna}
    \vspace{-0.4cm}
\end{figure*}

\vspace*{1ex}\noindent\textbf{Delayed Communication:}\label{sec:delayed}
A straightforward application of decentralized multi-agent approaches
like {\em independent Q-learning} (IQL)~\cite{Tan1997} to the Options
framework~\cite{sutton1999between} would yield agents that make
decisions independent of each other.  Agent $j$ would estimate the
Q-value (see Equation \ref{eq:options})
\begin{equation}
    Q^j_\text{iql}(s_t, o^j_t)
    := \E\!\Big[{\textstyle\sum\limits_{\tau=0}^{k-1}} 
        \gamma^\tau  r_{t+\tau} 
        \!+\! \gamma^k\!\! \max_{{o'}^j \in \mathcal{O}^j} \!\!Q^j_\text{iql}(s_{t+k}, o'^j)\Big] ,
\end{equation}
and select the option $o^j_t$ that maximizes it.  Here other agents
are treated as stationary parts of the environment, which can lead to
unstable training when those agents change their policy.  The best way
to avoid this instability would be learn the Q-value w.r.t.~the joint
option of all agents $\vec o_t := (o^1_t, \ldots, o^n_t)$,
i.e.~$Q_\text{joint}(s_t, \vec o_t)$
~\cite{riedmiller2005mutlioption}.  While these joint Q-values allow
training in a stationary environment, decisions require to maximize
over Q-values of all possible joint options.  As the number of joint
options grows exponentially in the number of agents $n$, and joint
optimization would require a vast communication overhead, this
approach is not feasible in decentralized scenarios.

Instead we propose to use a {\em delayed communication channel} 
over which agents signal the new option they switched to after each termination.
This reduces potentially costly communication and allows 
each agent $j$ access to all other agents' 
options of the previous time step 
$\vec o^{-j}_{t-1} := (o^1_{t-1}, \ldots, o^{j-1}_{t-1}, 
o^{j+1}_{t-1}, \ldots, o^n_{t-1})$.
Agents can approximate the joint Q-value 
by conditioning on this information, that is,
by choosing options $o^j_t$ that maximize 
the {\em delayed Q-value} $Q^j(s_t, \vec o^{-j}_{t-1}, o^j_t)$.
Note that the approximation is exact
if {\em no other agent terminates at time $t$}.
The optimality of the agents' decisions depends therefore 
on the frequency with which agents terminate their options.

\wbrm{
\section{Multi-agent Options Framework}
In this section, \wbrm{we will discuss the problems that we identified with existing multi-agent hierarchical model, and describe our methods that improves on these problems.}
Similarly to the extension to MAXQ framework proposed by \citeauthor{makar2001hierarchical}~\shortcite{makar2001hierarchical}, we extend the Options framework for the multi-agent model for a two-layer hierarchy. For agent $j$, the value \wbadd{$Q_\mu^j$} of choosing option $o^j$ at state $s_t$, while other agents are executing $o^{-j}$, is the total expected future return. \wbadd{This can be decomposed as the expected cumulative reward while executing $o = \{o^j, o^{-j}\}$, until the {\em first} agent's option $o^i$ terminates, that is, after $k = \min_i\big\{\argmin_k\{\beta(s_{t+k}, o^i)\}\big\}$ steps:}
\begin{align*}
    &Q_\mu^j(s_t,o^{-j},o^j) 
= \E\Big[\sum_{\tau=0}^\infty \gamma^\tau r_{t+\tau}
    \,\Big|\, o_t = o \Big] \\
& \qquad= \E\Big[\sum_{\tau=0}^{k-1} \gamma^\tau r_{t+\tau} \,\Big|\, o_t=o \Big] + \E\Big[ \sum_{\tau=k}^\infty \gamma^\tau r_{t+\tau} \Big]\\
& \qquad= R_{s_t}^o + \gamma^k \E\Big[ Q^j_\mu(s_{t+k}, o^{-j}_{t+k}, o^j_{t+k}) \Big] \,.
\end{align*}
The corresponding Bellman Optimality Equation is
\begin{align}
    \nonumber
    &Q^j_*(s_t, o^{-j}_t, o^j_t) \\
    &= \E\Big[\sum_{\tau=0}^{k-1} \gamma^\tau r_{t+\tau} + \gamma^k{\max_{o'\in \mathcal{O}} Q^j_*(s'_{t+k}, {o'}^{-j}, {o'}^j)}\Big] \,.
    \label{Bellman_optimal}
\end{align}
\wbadd{If all agents would react simultaneously to the first agent's termination, they would have to solve a centralized optimization problem over the joint option $o'_{t+k}$. To avoid the corresponding communication overhead, the agent must learn when the future communication costs outweigh the benefits of a quick reaction. To this end, we propose an algorithm based on multi-agent intra-option learning and dynamic termination.}
}

\vspace*{1ex}\noindent\textbf{Multi-agent Intra-Option Learning:}
As introduced in the previous section, 
the intra-option learning method (Equation \ref{eq:intra_option}) 
efficiently associates options with primitive actions. 
In our \wbadd{decentralized} multi-agent options model, 
agent $j$ selects an option according to $Q^j(s_t, \vec o^{-j}_{t-1}, o^j)$, 
which is defined as
\begin{eqnarray} \label{eq:our_intra_option}
    Q^j(s_t, \vec o^{-j}_{t-1}, o^j)
    & := &  \E\big[ r_t + \gamma 
        U^j(s_{t+1}, \vec o^{-j}_{t}, o^j) \big] \\
    \nonumber
    U^j(s_{t+1}, \vec o^{-j}_{t}, o^j)
    & := & \big(1 - \beta^{o^j}\!(s_{t+1}) \big) \;
        Q^j(s_{t+1}, \vec o^{-j}_{t}, o^j) \\
    \nonumber
    && + \, \beta^{o^j}\!(s_{t+1})  \!\max_{o'^j \in \mathcal{O}^j}\!\! 
        Q^j(s_{t+1}, \vec o^{-j}_{t}, o'^j)  . 
\end{eqnarray}
We can learn $Q^j$ by, for example, minimizing the mean-squared TD
error ~\cite{Sutton1998} between the left and right side of Equation
\ref{eq:our_intra_option}.  In line with intra-option learning, we
update the Q-values of all options $o^j$ that would have executed the
same action $a^j_t$ as the actually executed option $o^j_t$.  Note
that due to our delayed communication channel, the executed options of
all other agents 
are known after the transition to $s_{t+1}$ and can thus be used to
compute the target $U^j(s_{t+1}, \vec o^{-j}_{t}, o^j)$, that is, the
Q-value of either following the option $o^j$ if
$\beta^{o^j}\!(s_{t+1}) = 0$, or terminating and choosing another
option greedily if $\beta^{o^j}\!(s_{t+1}) = 1$.

\wbrm{
We extend Equation \ref{eq:intra_option} to the multi-agent case:
\begin{align*}
    \begin{split}
    &Q^j(s_t, o^{-j}_t, o^j) 
    \quad \leftarrow \quad 
    Q^j(s_t, o^{-j}_t, o^j) \\
    & \qquad + \alpha\, \big[r_{t+1} + \gamma U^j(s_{t+1}, o^{-j}_t, o^j) - Q^j(s_t, o_t^{-j}, o^j) \,\big] \,, \\[2mm]
    &U^j(s_{t+1},o_t^{-j}, o^j) := 
    (1-\beta(s_{t+1}, o^j)) \; Q^j(s_{t+1}, o_t^{-j} ,o^j) \\
    & \hspace{25mm} + \beta(s_{t+1}, o^j)\max_{o'^j\in \mathcal{O}^j}Q^j(s_{t+1}, o_t^{-j}, o'^j) \,.
    \end{split}
\end{align*}
}

\vspace*{1ex}\noindent\textbf{Dynamic Option Termination:}
As mentioned above, the delayed Q-value defined in Equation \ref{eq:our_intra_option}
only approximates the joint Q-value function.
This approximation will deteriorate when other agents terminate,
but sometimes agents can also benefit from early termination,
as shown in Figure \ref{fig:scene2}. 
Additionally, options are usually pre-trained 
and have to cover a large range of tasks, 
without being able to solve any one task perfectly.
Being able to prematurely terminate options 
can increase the expressiveness of the learned policy dramatically.

The easiest way to use partial options is to modify
the termination conditions $\beta^{o^j}(s)$.
In particular, we denote choosing the option with the 
largest Q-value (Eq.~\ref{eq:our_intra_option})
at each time step as {\em greedy termination}.
Following \cite{harutyunyan2017learning}
we combined this approach with an exploration
policy that terminates executed options 
with a fixed probability $\rho=0.5$
to allow for temporally extended exploration.
During testing the agent is nonetheless 
allowed to terminate greedily
at every step if the Q-value of another option is larger.

\wbrm{
The easiest way to use partial options is to modify
the termination conditions $\beta^{o^j}(s)$. 
We investigate the influence of different termination 
schemes on agents that maximize the delayed Q-value of Equation \ref{eq:our_intra_option}.
We also propose a new option termination method, 
which allows the agents' policies to learn when to terminate.\\

\noindent
\textbf{Greedy Termination} \quad
%
\citeauthor{harutyunyan2017learning}~\shortcite{harutyunyan2017learning} have shown that premature termination at each step 
to select the next best option can improve performance significantly, 
in particular in the presence of imperfect options.
They also show that this approach removes 
most exploration benefits of using options.
The advantage of sampling temporally extended options
can be somewhat reclaimed by using a separate {\em exploration policy},
which follows the current option and randomly terminates it with a fixed probability $\rho$
(in our experiments $\rho=0.5$).
However, during testing the agent is allowed to terminate greedily
at every step if the Q-value of another option is larger.\\
}


\wbrm{\noindent \textbf{Dynamic Termination} \quad} Although greedy
termination has been shown to improve the performance of individual
agents with imperfect options \cite{harutyunyan2017learning}, the
agent's behaviour will become less predictable for others.  In
particular, agents that utilize the delayed Q-value of Equation
\ref{eq:our_intra_option} will make sub-optimal decisions whenever
another agent terminates.  To increase the predictability of agents,
while allowing them to terminate flexibly when the task demands it, we
propose to put a price $\delta$ on the decision to terminate the
current option.  Option termination is therefore no longer hard-coded,
but becomes part of the agent's policy, which we call {\em dynamic
  termination}.  This can be represented by an additional option
$o^j=T$ for agent $j$ to terminate.  Note that, unlike in the Options
framework, we no longer need a termination function
$\beta^{o^j}\!(s_t)$ for each option $o^j$.  It is sufficient to
compare the value of the previous option
$Q^j(s_t, \vec o^{-j}_{t-1}, o^j_{t-1})$ with the value of termination
$Q^j(s_t, \vec o^{-j}_{t-1}, T)$.  Evaluating $o^j=T$ is
computationally similar to evaluating the termination condition
$\beta^o$.  Dynamic termination therefore has a similar cost to
traditional termination.

The optimal behaviour for a given punishment $\delta$
is the fix-point of the novel {\em dynamic termination Bellman equation}:
\wbrm{
\begin{equation}\label{dyna}
    Q^j(s_t, \vec o_{t-1}^{-j}, o^j) := \kern-0.5ex\left\{\kern-1ex
    \begin{array}{ll}
        \E\big[r_t \!+ \gamma \kern-2ex\max\limits_{o'^j \in \{o^j, T\}}\kern-2ex 
            {Q^j(s_{t+1}, \vec o_{t}^{-j}\kern-1ex, o'^j)\big]}\kern-2ex 
            &, o^j \!\neq\! T \\
        \quad\; \max\limits_{o'^j \in \mathcal{O}^j}{Q^j(s_t, \vec o_{t-1}^{-j}, o'^j)} 
            \,-\, \delta \! &, o^j \!=\! T,
    \end{array}
    \right. \!\!
\end{equation}
}

\vspace{-4mm}
\begin{eqnarray}
    \nonumber
        Q^j(s_t, \vec o_{t-1}^{-j}, o^j \!\neq\! T) &\kern-1.5ex:=\kern-1.5ex&
            \E\big[r_t \!+ \gamma 
                \max\limits_{\kern-1ex o'^j \in \{o^j, T\}}
                {Q^j(s_{t+1}, \vec o_{t}^{-j}\kern-1ex, o'^j)\big]} , 
    \\ \label{dyna}
        Q^j(s_t, \vec o_{t-1}^{-j}, o^j \!=\! T) & := & 
            \max\limits_{o'^j \in \mathcal{O}^j}{Q^j(s_t, \vec o_{t-1}^{-j}, o'^j)} \,-\, \delta \,.
\end{eqnarray}
Similarly to Equation \ref{eq:our_intra_option}, Equation \ref{dyna}
allows intra-option learning and can be applied to all options $o^j$
that would have selected the same action as the executed option
$o^j_t$.  Note that the termination option $T$ can always be updated,
as it does not depend on the transition.

\wbrm{
We will use the most straight-forward model for illustrating the idea: with two separate policies: the option selection policy $\mu_o$, and the termination policy $\mu_\beta$.

For the termination policy $\mu_\beta$, The value of choosing to continue ($\beta = 0$) with option $o$ at $s_t$ can be approximated by the one-step intra-option learning target, and instead of using the target value $U(s_{t+1}, o)$ in Equation.\ref{eq:intra_option}, which is determined by the termination condition defined by the options, we use the termination condition chosen by the termination operator at $s_{t+1}$, assuming the agent will choose optimally to terminate or continue with $o$ at the next step $s_{t+1}$.

On the other hand, if the option $o$ is terminated at $s_t$, the agent will choose and execute the optimal option according to the option selection policy $\mu_o$. Therefore, the value of terminating an option $o$ at $s_t$ will be the value of continuing with the optimal option at $s_t$, and in addition, a penalty $\delta$, which can be the penalty for switching options. $\delta$ can be either a hand-tuned value for balancing the control performance and switch frequency, or a value defined by the environment, eg. a driver can be penalized for constantly switching directions, which is dangerous for the other cars.

The model can be simplified based on the following observations:
1. The value of selecting option $o$ according to $\mu_o$, is the same as the value of choosing to continue with $o$ by $\mu_\beta$. 2. For $\mu_\beta$, the value of terminating any option will be the same: which is the value of continuing with the optimal policy added with the switching penalty.

Therefore, both $\mu_o$ and $\mu_\beta$ can be modeled jointly by the option termination Q network $\mu_\beta$, which outputs the value of continuing with each option and a single entry for the value of terminating any option at a given state, as shown in Figure.\ref{fig:dyna}.}

\vspace*{1ex}\noindent\textbf{Deep Q-learning:}
A group of $n$ agents can be trained using 
a deep Q-network $Q_\theta$~\cite{Mnih2013}, 
parameterized by $\theta$.
The architecture is shown in Figure \ref{fig:dyna}:
each agent $j$ selects and executes the next option 
based on the current state (i.e.~grid-word image) $s_t$ 
and the last known options $\vec o^{-j}_{t-1}$ of all the other agents. 
A centralized manager is not needed, 
and the options must only be broadcast 
after an agent chose to select a new option. 
The Q-value of choosing an option is updated 
by {\em temporal difference learning} with {\em experience replay},
which is the established standard procedure 
in deep Q-learning~\cite{Mnih2013}. 
To reduce the number of parameters, 
we let all the agents share $\theta$, 
and the model is thus updated using the experiences 
collected by all the agents.
To differentiate the behaviour of different agents,
the presented grid-world image contains a dedicated 
channel that encodes the current agent's state.
These design decisions follow previous work 
in deep multi-agent learning 
\cite{foerster2017counterfactual,rashid2018qmix}
and drastically reduce training time 
with very little impact on the performance in large domains.

At each transition $t$, the Q-values of all options $o^j$, that would execute the same action as the executed option $o_t^j$, and the termination option $T$ are updated by gradient descent on the sum of their respective losses
\begin{eqnarray}
    \nonumber
    \mathcal{L}^{o}_{j,t}\big[\theta\big] 
    & := & \Big( 
        r_t \!+\! \gamma \max_{o' \in \{o, T\}}
            Q_{\theta}(s_{t+1}, \vec o^{\!-j}_{t}\!\!, o')
        - Q_\theta(s_t, \vec o^{-j}_{t-1}, o) \Big)^{\!2}, 
\\[-1mm]
    \nonumber
    \mathcal{L}^T_{j,t}\big[\theta\big] 
    & := & \Big(
        \max\limits_{o' \in \mathcal{O}^j}
            Q_{\theta}(s_t, \vec o_{t-1}^{-j}, o')
        - \delta 
        - Q_\theta(s_t, \vec o_{t-1}^{-j}, T) \Big)^{\!2} \kern-.5ex.
\end{eqnarray}
The total loss for a batch of $m$ transitions with $n$ agents is 
\vspace{-1mm}
\begin{equation} 
    \mathcal{L}\big[\theta\big] \quad:=\quad 
    \frac{1}{mn}
    \sum_{t=0}^{m-1} \sum_{j=1}^n 
    \Big( 
    \mathcal{L}^T_{j,t}\big[\theta\big] \;
    + \kern-2ex\sum_{\pi^o(s_t) = a^j_t}\kern-2ex
    \mathcal{L}^{o}_{j,t}\big[\theta\big] \Big)\,.
\end{equation}
\vspace{-1cm}

%% file: experiments.tex
\begin{table*}[t]
\hspace{-0.5cm}
\begin{minipage}[c][4cm][c]{\textwidth}
    \begin{minipage}[b][4cm][b]{0.48\textwidth}
    \captionsetup{type=figure}
    \centering
    \includegraphics[width=6.4cm, height = 4.3cm]{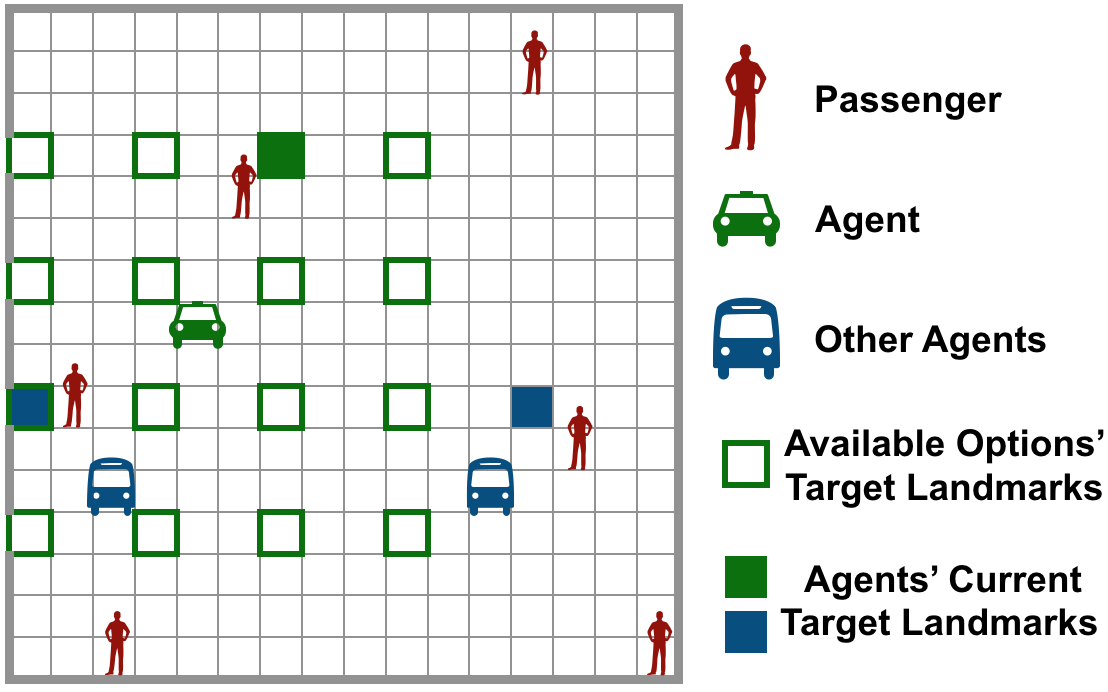}
    \caption{Example $16\times 16$ Grid-world}
    \vspace{0.5cm}
    \label{fig:demo}
    \end{minipage}
    \hspace{0.6cm}
    \begin{minipage}[b][7cm][b]{0.6\textwidth}
    \vspace{0pt}
    \begin{center}
    \renewcommand{\arraystretch}{1.0}
    \input{./table_flexibility_predictability.tex}
    \end{center}
    \vspace{-5mm}
    \end{minipage}
\end{minipage}
\vspace{1cm}
\end{table*}

We will first evaluate the flexibility and predictability of our dynamically terminating agent, followed by the impact of dynamic termination on the agents' performance.

\vspace*{1ex}\noindent\textbf{Experimental Setup:}
Figure \ref{fig:demo} shows a $16\times 16$ grid-world of the taxi
pickup as observed by the green agent, which includes the passengers,
the other agents and their broadcast options. The landmarks of
distance $L=3$ show the destinations of options that are currently
visible to the agent. This raises our \textit{first challenge}: in
order to reach a passenger that stands outside the landmarks, an agent
needs to correctly switch between options.

In the \textit{Taxi Pickup Task} $m$ passengers are randomly
distributed in each episode. An agent is rewarded $r=1$ when occupying
the same grid as a passenger, and each step incurs a cost of
$-0.01$. Apart from landmark switching, the agents need to interpret
others' behaviours to avoid choosing the same passenger, as well as
responding quickly to changes such as when a passenger is picked up by
another agent.

In the \textit{Pursuit Task} agents try to catch randomly distributed
prey by cooperating with others. We refer to the task as $k$-agent
pursuit, where a successful capture requires at least $k$ agents
occupying $k$ positions adjacent to the prey, which rewards each
participating agent $r =1$. This task relies heavily on agents
coordination. In particular, when close to a specific prey, agents
need to observe others and switch between options to surround the
prey; whereas when faraway, agents need to agree on and commit to go
for the same prey.

\vspace*{1ex}\noindent\textbf{Algorithms and Training:} Having
described the settings, we now introduce the detailed training
procedures of the SMDP and option policies, before comparing the four
types of agents.

\textit{The Policy of Options} adopts a local perspective, and navigates the agent to the option's destination. Specifically, we use a DQN of $2$ convolutional layers (kernel size 2) with max-pooling, followed by $4$ fully-connected layers (size 300). The input is the destination coordinate with the grid-world image, and the output is a primitive action in \{N, S, E, W, Stay\}.

\dgrm{For example, an agent at position ($x_a$, $y_a$), choosing an option with the sub-goal ($x_g$, $y_g$), will effectively execute the option with sub-goal ($x_g - x_a$, $y_g - y_a$). }

\textit{The SMDP Policies} are trained through intra-option learning for all agent types. The inputs are $4$-channel grid-world images as in Figure \ref{fig:dyna}, which represents the agent, the preys (or passengers), the other agents, and lastly, the options broadcast by other agents (except for IQL). The DQN contains 2 convolutional layers (kernel size 3), max pooling, and 4 fully-connected layers (size 512). We use experience replay with a replay buffer of size 100,000.

\textit{Self-Play}
is used in the experiments, and our decentralized agents share the same DQN parameters (not states)~\cite{rashid2018qmix}. This allows us to scale up the number of agents without additional parameters; and the trained model can directly transfer to more agents during testing. Moreover, self-play creates an important link between the predictability of an individual agent and of the society.

\noindent
The four types of agents are as follows:
\begin{enumerate}
\item \textit{Option Termination Agent} executes its option until the natural termination condition is met.
\item \textit{Greedy Termination Agent} terminates every step and
  switches to the optimal option. For better exploration during
  training, an additional behaviour policy is used for experience
  collection. For fairness of comparison, this exploration policy
  which terminates with probability $\rho=0.5$ (tuned for the greedy
  agent) is applied across all agent types.
\item \textit{Dynamic Termination Agent} is our proposed algorithm
  that chooses whether to terminate the current option at each
  step.\dgrm{, by comparing the Q-value of this option and of the
    termination option $T$. On termination, the agent will switch to
    the optimal option.} $\delta$ is the termination penalty.
\item \textit{IQL} is independent Q-learning, where agents option
  broadcasts are disabled. IQL (greedy) and IQL ($\delta$) refers to
  IQL agents using greedy or dynamic termination.
\end{enumerate}

\begin{figure*}[t]
 \captionsetup{type=figure}
  \subfigure[$19\times 19$ taxi with 10 agents]{\label{fig:19_taxi}\includegraphics[width=0.5\textwidth]{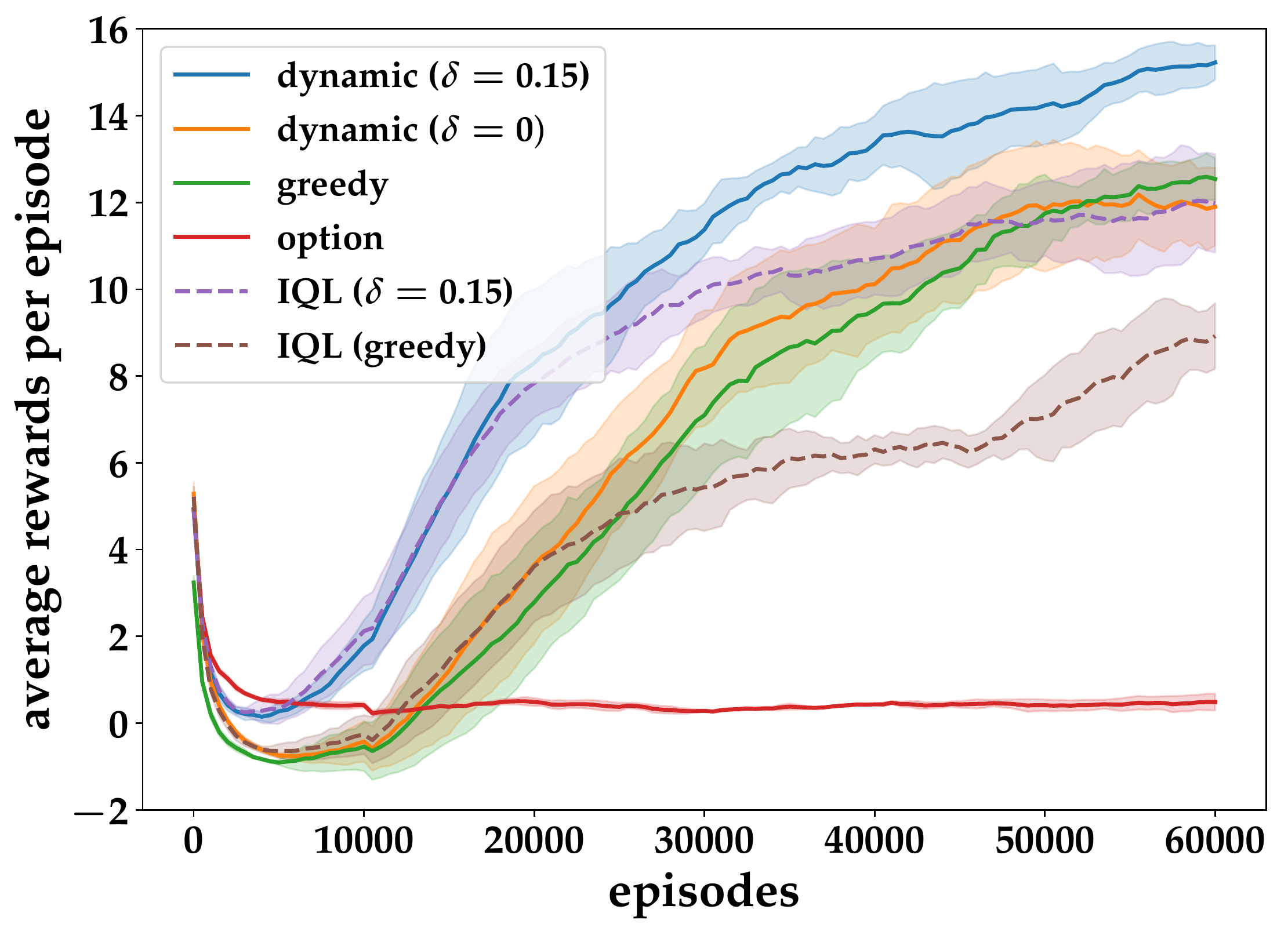}}
  \hspace{-0.2cm}
  \subfigure[$19\times 19$ pursuit with 10 agents]{\label{fig:pursuit}\includegraphics[width=0.5\textwidth]{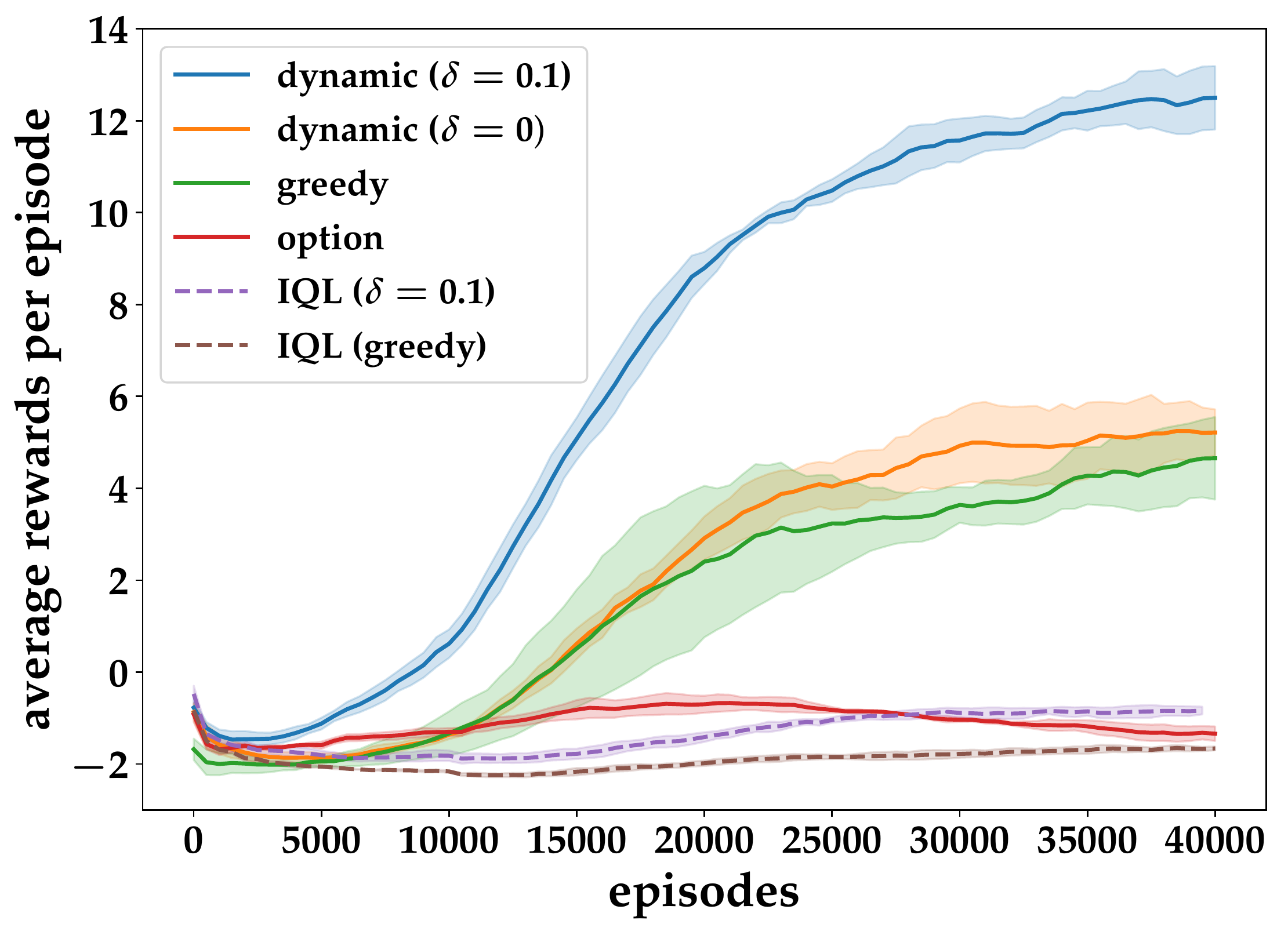}}
  \vspace{-0.4cm}
  \caption{Results from Taxi Pickup and Pursuit Tasks}
     \small{Note: Every point per 500 episode is the testing result averaged over 100 random episodes and 5 seeds. The shaded area shows standard deviation across seeds.}
\end{figure*}

\begin{table*}[!t]
\begin{center}
\renewcommand{\arraystretch}{1.0}
\input{table.tex}
\end{center}
\vspace{-1cm}
 \end{table*}

\vspace*{1ex}\noindent\textbf{Results:} The delayed response problem
reveals that agents need to be flexible enough to change their options
when the situation changes, but also predictable enough not to
interfere in other agents' plans too frequently.  Table \ref{table:
  flexibility} shows experimental measurements to showcase these
conflicting goals for the investigated termination methods.
We measure the agents' {\em flexibility} in the single-agent taxi domain.  
100 episodes are initialized with 5 random passengers. 
During each episode, 
one additional passenger is placed near the agent at step $T$ 
and we observe how quickly the agent adjusts to the new situation. 
We report the percentage of option changes at step $T+1$ 
and the average number of steps till the agent changes options.
Note that dynamic termination allows to react 
almost as flexible to the changed situation as the greedy termination.

For {\em predictability}, 
we measure the average probability to change the option 
in the multi-agent taxi task for two cases: 
when the agent is near (within distance 4) 
or far from its closest passenger. 
This is an imperfect measurement, 
as we cannot distinguish the effect of termination on other agents.
While options need to change close to a passenger
due to imperfect options,
the behavior of dynamic termination is much closer
to standard option termination when far away.
Note that this effect is marginal for the other techniques,
which indicates that our method may purposefully 
refrain from changing to better options 
to avoid interrupting other agent's plans.


\textit{Performance:}
Figure \ref{fig:19_taxi} shows the results from the taxi pickup task. The option termination agent fails due to its inflexibility to switch options. In contrast, our dynamic ($\delta=0.15$) agent is highly flexible. Moreover compared with greedy and IQL, its high predictability indeed helps the agents to interpret others' intentions and better distribute their target passengers.
Figure \ref{fig:pursuit} shows the results on the pursuit task, where at least two agents need to surround a prey within capture range $= 1$. Seen from the IQL agents' low performance, option broadcasting and interpreting others' behaviours are crucial to this task. Our dynamic termination agent ($\delta=0.1$) significantly outperforms all other agents. Compared with the greedy agents, we can conclude that predictability significantly helps our dynamic agents to stay committed and succeed in cooperation.

\wbrm{
\textit{The $\delta$ Penalty:}
Figure \ref{fig:term_taxi} and \ref{fig:term_pursuit} show the average number of terminations of each agent per episode. Note that greedy and IQL(greedy) incurs significant terminations, while our dynamic termination agent with $\delta=0.15$ and $\delta=0.1$ significantly reduce the number of terminations.
In Figure \ref{fig:pursuit_delta} and \ref{fig:taxi_delta} we plot the results of our dynamic terminating agent trained across varying $\delta$ penalty for switching. Observe that: 1. from Figure \ref{fig:19_taxi} and \ref{fig:pursuit}, the result of $\delta = 0$ roughly overlaps with that of the greedy agent.
2. Comparing the $\delta=1$ with $\delta=0$, we observe a significant increase in the rewards, especially in the pursuit task where coordination is crucial.
3. As $\delta$ increases from the optimum, the curve stays stable. This demonstrates the robustness of our dynamic terminating agent against perturbations of $\delta$.
}

Finally, we present the performance of all agents across different
tasks and varying parameters in Table \ref{table: all}. Firstly, the
option termination agent has difficulty with tasks which require
higher level of accuracy and quick responses, such as the taxi tasks
and pursuit with capture range $=1$. However, it works well with tasks
which require coordination but less flexibility, such as the
$16\times 16$ 3 agent pursuit with capture range 2, which shows the
advantage of predictability on cooperation. The performance of greedy
termination agents decreases significantly in larger grid-world sizes,
and when commitment is essential, such as the $16\times 16$ 3 agent
pursuit with capture range 2. Our dynamically terminating agent
performs well across all tasks, as it balances well between
flexibility and predictability. The IQL agents performs well in the
taxi task. However, they fail to learn the pursuit tasks where
foreseeing others' behaviours is essential to coordination.

%% file: table_flexibility_predictability.tex
\scriptsize
\begin{tabular}{@{}cccccc@{}}
\toprule
 & \multicolumn{3}{c}{\textbf{predictability}} & \multicolumn{2}{c}{\textbf{flexibility}} \\ \midrule
 & \multicolumn{3}{c}{\textbf{\begin{tabular}[c]{@{}c@{}}option\\ changes\end{tabular}}} & \textbf{\begin{tabular}[c]{@{}c@{}}option\\ changes\end{tabular}} & \textbf{\begin{tabular}[c]{@{}c@{}}steps to\\ change\end{tabular}} \\ \midrule
 & \textit{\textbf{all}} & \textit{\textbf{near}} & \textit{\textbf{far}} & \textit{\textbf{all}} & \textit{\textbf{all}} \\ \midrule
\textbf{dynamic} & 24.1\% & 28.1\% & 16.8\% & 63\% & 1.61 \\ \midrule
\textbf{greedy} & { \textbf{59.9\%}} & { \textbf{57.9\%}} & { \textbf{52.8\%}} & { \textbf{77\%}} & 1.15 \\ \midrule
\textbf{option} & 10.9\% & 10.9\% & 9.5\% & 3\% & 6.86 \\ \bottomrule
\end{tabular}
\vspace{0.1cm}
\caption{Flexibility and Predictability Results. Near/far refers to whether an agent is within distance $=4$ to a passenger. Steps to change denotes the number of steps from the new passenger is placed to the agent's option change. \dgrm{The results are averaged over 100 random episodes across 3 seeds}}
\label{table: flexibility}

%% file: table.tex
\scriptsize\setlength{\tabcolsep}{1pt}
\begin{tabular}{@{}cccccccccc@{}}
\toprule
\multicolumn{2}{c}{} & \multicolumn{3}{c}{\textbf{taxi}} & \multicolumn{3}{c}{\textbf{2 agent pursuit}} & \multicolumn{2}{c}{\textbf{3 agent pursuit}} \\ \midrule
 &  & \multicolumn{2}{c}{\textbf{n=5, m=10}} & \textbf{n=10, m=20} & \multicolumn{2}{c}{\textbf{n=3, m=5}} & \textbf{n=10, m=10} & \multicolumn{2}{c}{\textbf{n=3, m=5}} \\ \midrule
\textbf{Agents} &  & \textit{\textbf{19x19}} & \textit{\textbf{25x25}} & \textit{\textbf{19x19}} & \textit{\textbf{\begin{tabular}[c]{@{}c@{}}16x16\\ (r=1)\end{tabular}}} & \textit{\textbf{\begin{tabular}[c]{@{}c@{}}19x19\\ (r=1)\end{tabular}}} & \textit{\textbf{\begin{tabular}[c]{@{}c@{}}19x19\\ (r=1)\end{tabular}}} & \textit{\textbf{\begin{tabular}[c]{@{}c@{}}10x10 \\ (r=1)\end{tabular}}} & \textit{\textbf{\begin{tabular}[c]{@{}c@{}}16x16\\  (r=2)\end{tabular}}} \\ \midrule
\multirow{2}{*}{\textbf{Dynamic}} & \textit{$\delta$ = 0.1} & \textbf{7.89} & \textbf{5.75} & \textbf{15.29} & \textbf{10.24} & \textbf{9.30} & \textbf{12.50} & \textbf{6.71} & \textbf{10.38} \\ \cmidrule(l){2-10} 
 & \textit{$\delta$ = 0} & 6.58 & 3.28 & 11.81 & 6.73 & 4.07 & 5.38 & 5.53 & 6.54 \\ \midrule
\textbf{Greedy} & \textit{} & 6.62 & 3.23 & 12.39 & 7.36 & 3.74 & 4.65 & 5.89 & 6.40 \\ \midrule
\textbf{Option} & \textit{} & -0.32 & -0.94 & 0.52 & 5.47 & -1.82 & -1.42 & -3.77 & 5.20 \\ \midrule
\multirow{2}{*}{\textbf{IQL}} & \textit{$\delta$ = 0.1} & 7.11 & 5.09 & 12.02 & -1.57 & -2.29 & -0.84 & -1.62 & -0.59 \\ \cmidrule(l){2-10} 
 & \textit{greedy} & 6.08 & 2.79 & 9.06 & -2.12 & -2.49 & -1.64 & -2.13 & -0.42 \\ \bottomrule
\end{tabular}
\centering
\vspace{0.1cm}
\caption{Average reward after training for Taxi and Pursuit tasks. n is the number of agents and m is the number of passengers (preys). NxN denotes grid-world size, k agent pursuit denotes the required number of agents for capture, and r is the capture range.}
\label{table: all}

%% file: conclusion.tex
In this paper, we identified the delayed response problem, that occurs when hierarchical RL is combined with multi-agent learning. To address this challenge, we investigated existing approaches of greedy option termination in single agent learning. However, this method introduces a new dilemma specific to multi-agent systems: as an agent broadcasts its current options to indicate its subsequent behaviours, frequent changes in options will result in its behaviour being less predictable by others. Therefore, to balance flexibility with predictability, we introduced dynamic termination, which enables agents to terminate their options flexibly according to the current state. We compared our model with current state of the art algorithms on multi-agent pursuit and taxi tasks with varying task parameters, and demonstrated that our approach outperformed the baselines through flexibly adapting to the task requirements. For future work, we are interested in applying the dynamic termination framework to traffic simulations, such as junction and highway management. \dgrm{And we are interested in  broadcasting the potential termination behavior to equip the agents with better estimation of others' commitment to their broadcasted options.}